\documentclass[conference, a4paper]{IEEEtran}
\usepackage{algorithm,algorithmic}
\usepackage{amsmath,amssymb,mathrsfs}
\usepackage{array}
\usepackage{balance}
\usepackage{booktabs}
\usepackage{cite}
\usepackage{epsfig}
\usepackage{fancyhdr}
\usepackage{float}
\usepackage[nomain,acronym,automake,toc,nopostdot]{glossaries}
\usepackage{graphicx,color}
\usepackage{mathtools}
\usepackage{multirow}
\usepackage{psfrag}
\usepackage{stfloats}
\usepackage{subfigure}
\newtheorem{lem}{Lemma}

\newtheorem{thm}{Theorem}

\makeglossaries
\newacronym{1d}{1D}{one-dimensional}
\newacronym{2d}{2D}{two-dimensional }
\newacronym{3d}{3D}{three-dimensional}
\newacronym{3gpp}{3GPP}{3rd Generation Partnership Project}
\newacronym{5g}{5G}{fifth-generation}
\newacronym{6g}{6G}{sixth-generation}

\newacronym{ap}{AP}{access point}
\newacronym{amp}{AMP}{approximate message passing}
\newacronym{ask}{ASK}{amplitude shift keying}
\newacronym{awgn}{AWGN}{additive white Gaussian noise}

\newacronym{ber}{BER}{bit error rate}
\newacronym{bfgs}{BFGS}{Broyden–Fletcher–Goldfarb–Shanno}
\newacronym{bler}{BLER}{bit error rate}
\newacronym{bp}{BP}{belief propagation}
\newacronym{bpsk}{BPSK}{binary phase shift keying}
\newacronym{bs}{BS}{base station}

\newacronym{cbsm}{CBSM}{correlation-based stochastic model}
\newacronym{csirs}{CSI-RS}{channel state information reference signal}
\newacronym{cdf}{CDF}{cumulative distribution function}
\newacronym{cdma}{CDMA}{code division multiple access}
\newacronym{cg}{CG}{conjugate gradient descent}
\newacronym{clt}{CLT}{central limit theorem}
\newacronym{csi}{CSI}{channel state information }
\newacronym{cvp}{CVP}{closest vector problem}

\newacronym{dof}{DoF}{degrees of freedom}

\newacronym{edw}{EDW}{exponentially decaying window}
\newacronym{elaa}{ELAA}{extremely large aperture array}
\newacronym{etsi}{ETSI}{European Telecommunications Standards Institute}

\newacronym{ff}{FF}{far-field}
\newacronym{fcsd}{FCSD}{fixed-complexity sphere decoder}
\newacronym{fec}{FEC}{forward error correction}
\newacronym{fspl}{FSPL}{free space path loss}

\newacronym{gbsm}{GBSM}{geometry-based stochastic model}
\newacronym{gd}{GD}{gradient descent}
\newacronym{gmsk}{GMSK}{Gaussian minimum shift keying}
\newacronym{gs}{GS}{Gauss-Seidel}
\newacronym{gsm}{GSM}{global system for mobile communication}

\newacronym{iid}{i.i.d.}{independently and identical distributed}
\newacronym{ils}{ILS}{integer least-squares}
\newacronym{ind}{i.n.d.}{independently and non-identical distributed}
\newacronym{imt}{IMT}{International Mobile Telecommunications}
\newacronym{isac}{ISAC}{integrated sensing and communication}
\newacronym{isi}{ISI}{intersymbol interference}
\newacronym{itur}{ITU-R}{International Telecommunication Union Radiocommunication Sector}
\newacronym{iui}{IUI}{inter-user interference}

\newacronym{ji}{JI}{Jacobi iteration}
\newacronym{jsac}{JSAC}{joint sensing and communication}

\newacronym{las}{LAS}{likelihood ascent search}
\newacronym{lbfgs}{LBFGS}{limited-memory Broyden–Fletcher–Goldfarb–Shanno}
\newacronym{leo}{LEO}{low Earth orbit}
\newacronym{lll}{LLL}{Lenstra-Lenstra-Lov\'{a}sz}
\newacronym{llr}{LLR}{log-likelihood ratio}
\newacronym{los}{LoS}{line of sight}
\newacronym{lr}{LR}{lattice reduction}
\newacronym{lmmse}{LMMSE}{minimum mean square error}
\newacronym{ls}{LS}{least square}
\newacronym{lsd}{LSD}{list sphere decoder}
\newacronym{lte}{LTE}{long-term evolution}

\newacronym{map}{MAP}{maximum a posteriori}
\newacronym{mf}{MF}{matched filter}
\newacronym{mimo}{MIMO}{multiple-input multiple-output}
\newacronym{mld}{MLD}{maximum likelihood detection}
\newacronym{mmimo}{mMIMO}{massive multiple-input multiple-output}
\newacronym{mrc}{MRC}{maximum ratio combining}
\newacronym{mse}{MSE}{mean square error}
\newacronym{ms}{MS}{matrix-splitting}
\newacronym{mt}{MT}{mobile terminal}
\newacronym{nf}{NF}{near-field}
\newacronym{nlos}{NLoS}{non-LoS}

\newacronym{od}{OD}{orthogonality defect}

\newacronym{pdf}{PDF}{probability distribution function}
\newacronym{pda}{PDA}{probabilistic data association}
\newacronym{pep}{PEP}{pairwise error probability}
\newacronym{pl}{PL}{path-loss}
\newacronym{pmf}{PMF}{probability mass function}
\newacronym{pwm}{PWM}{plane-wave model}

\newacronym{qam}{QAM}{quadrature amplitude modulation}
\newacronym{qpsk}{QPSK}{quadrature phase shift keying}
\newacronym{qn}{QN}{quasi-Newton}
\newacronym{rts}{RTS}{reactive tabu search}
\newacronym{ri}{RI}{Richardson iteration}
\newacronym{ris}{RIS}{reconfigurable intelligent surface}
\newacronym{rss}{RSS}{received signal strength}
\newacronym{rzf}{RZF}{regularized-ZF}

\newacronym{sa}{SA}{Seysen's algorithm}
\newacronym{sd}{SD}{steepest descent}
\newacronym{sdr}{SDR}{semidefinite relaxation}
\newacronym{ser}{SER}{symbol error rate}
\newacronym{sf}{SF}{shadow fading}
\newacronym{sic}{SIC}{succesive interference cancellation}
\newacronym{sinr}{SINR}{signal-to-interference-plus-noise ratio}
\newacronym{siso}{SISO}{single input single output}
\newacronym{snr}{SNR}{signal-to-noise ratio}
\newacronym{sns}{SNS}{spatial non-stationarity}
\newacronym{sota}{SoTA}{state-of-the-art}
\newacronym{ssor}{SSOR}{symmetric successive over-relaxation}
\newacronym{svd}{SVD}{singular value decomposition }
\newacronym{swm}{SWM}{spherical-wave model}

\newacronym{tr}{TR}{Technical Report}
\newacronym{ts}{TS}{tabu search}

\newacronym{uca}{UCA}{uniform cylindrical array}
\newacronym{ue}{UE}{user equipment}
\newacronym{ula}{ULA}{uniform linear array}
\newacronym{ura}{URA}{uniform rectangular array}
\newacronym{uma}{UMa}{urban macro}
\newacronym{umi}{UMi}{urban micro}
\newacronym{upa}{UPA}{uniform planar array}
\newacronym{ut}{UT}{user terminal}

\newacronym{vblast}{V-BLAST}{vertical Bell Labs layered space-time}

\newacronym{xlmimo}{XL-MIMO}{extra-large multiple-input multiple-output}

\newacronym{zf}{ZF}{zero-forcing}

\definecolor{sblue}{RGB}{0,51,120}
\definecolor{sred}{RGB}{200,51,130}

\newcommand{\figref}[1]{Fig. \ref{#1}}

\newcommand{\secref}[1]{Section \ref{#1}}
\newcommand{\thmref}[1]{{\it Theorem \ref{#1}}}
\renewcommand{\eqref}[1]{(\ref{#1})}

\ifCLASSINFOpdf
\else
\fi

\begin{document}
\title{Cluster-Aware Two-Stage Method for Fast Iterative MIMO Detection in LEO Satellite Communications}
\author{Jiuyu Liu, Yi Ma$^{\dagger}$, Qihao Peng, and Rahim Tafazolli\\
	{\small 5GIC and 6GIC, Institute for Communication Systems, University of Surrey, Guildford, UK, GU2 7XH}\\
	{\small Emails: (jiuyu.liu, y.ma, q.peng, r.tafazolli)@surrey.ac.uk}}
\markboth{}%
{}

\maketitle

\begin{abstract}
In this paper, a cluster-aware two-stage multiple-input multiple-output (MIMO) detection method is proposed for direct-to-cell satellite communications.
The method achieves computational efficiency by exploiting a distinctive property of satellite MIMO channels: users within the same geographical cluster exhibit highly correlated channel characteristics due to their physical proximity, which typically impedes convergence in conventional iterative MIMO detectors.
The proposed method implements a two-stage strategy that first eliminates intra-cluster interference using computationally efficient small matrix inversions, then utilizes these pre-computed matrices to accelerate standard iterative MIMO detectors such as Gauss-Seidel (GS) and symmetric successive over-relaxation (SSOR) for effective inter-cluster interference cancellation.
Computer simulations demonstrate that the proposed method achieves more than $12$ times faster convergence under perfect channel state information.
Even when accounting for channel estimation errors, the method maintains $9$ times faster convergence, demonstrating its robustness and effectiveness for next-generation satellite MIMO communications.
\end{abstract}

\begin{IEEEkeywords}
	MIMO detection, satellite communications, fast iterative methods, interference cancellation.
\end{IEEEkeywords}

\section{Introduction}\label{sec1}
Direct-to-cell satellite communication represents a critical frontier for next-generation global connectivity, offering unprecedented coverage capabilities beyond the limitations of traditional terrestrial infrastructure \cite{ITUR2023}.
The deployment of \gls{mimo} antenna arrays on \gls{leo} satellites enables simultaneous service to multiple users within identical time-frequency bands, substantially enhancing area spectral efficiency \cite{3GPP2023}.
While theoretically optimal, the \gls{mld} approach encounters exponential computational complexity that becomes impractical even with a modest number of users \cite{Liu2023}.
Linear detection methods \gls{lmmse} provide enhanced computational efficiency but remain constrained by cubic-order complexity due to matrix inversion requirements, which cannot be easily parallelized \cite{Albreem2019}.
In terrestrial massive-\gls{mimo} systems, the channel matrix typically exhibits well-conditioned properties, a characteristic known as favorable propagation \cite{Ngo2014}.
Under these conditions, iterative methods including \gls{ri}, \gls{gs}, \gls{ssor}, et al., can circumvent matrix inversions while achieving quadratic complexity and rapid convergence \cite{Zhang2021, Xie2016, Liu2024b, Tu2020}.

\begin{figure}[t]
	\centering
	\begin{minipage}[t]{0.49\textwidth}	
		\centering
		\includegraphics[width=8.5cm]{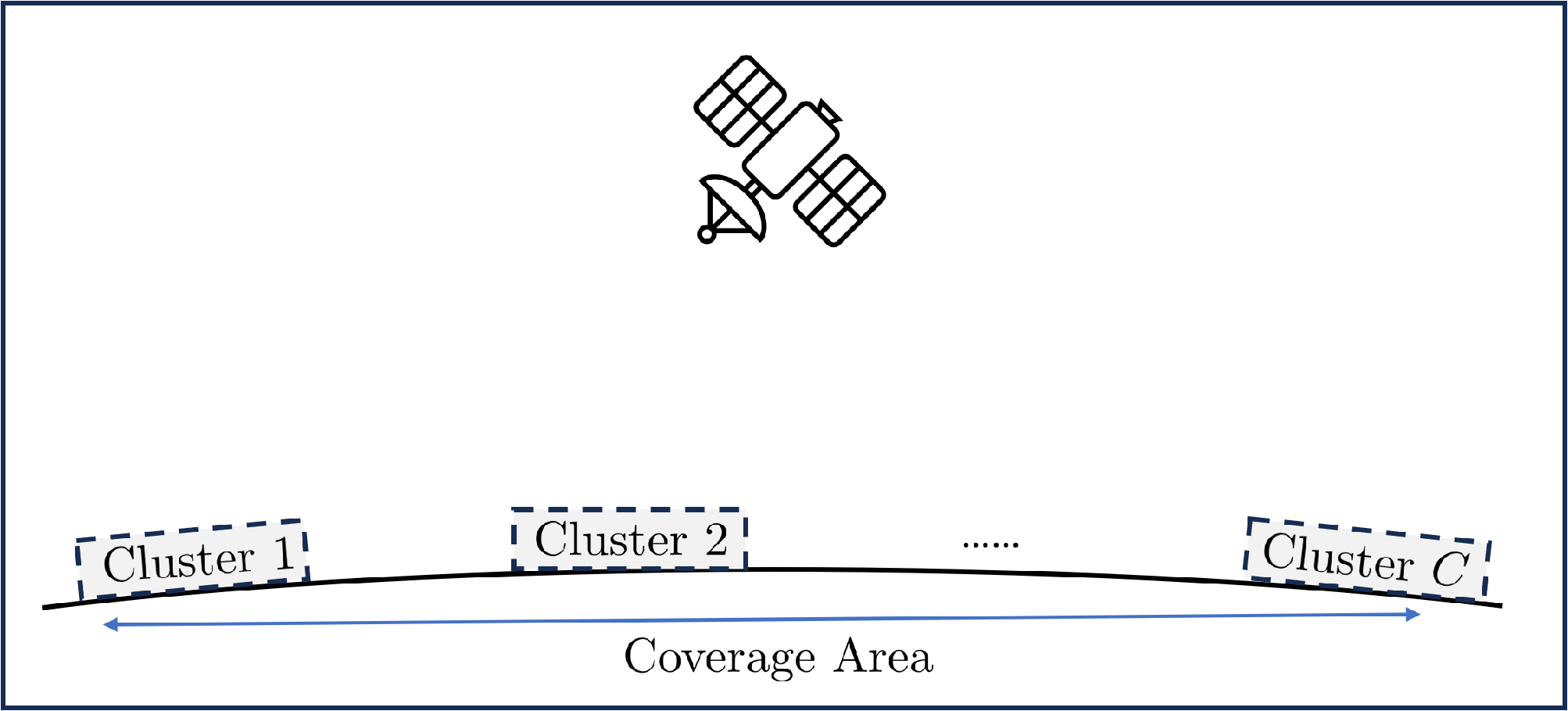}
	\end{minipage}
	\caption{\label{fig01} Non-uniform user distribution in the coverage area of a satellite. Users maintain close proximity within clusters while substantial distances separate different clusters, creating varying degrees of channel correlation.}
	\vspace{-1em}
\end{figure}  
However, when directly applying these iterative methods for satellite MIMO detection, significant convergence challenges emerge due to channel ill-conditioning.
This primarily occurs because satellite MIMO channels operate predominantly in \gls{los} conditions with high spatial correlation \cite{Xiang2024}.
In practice, the coverage of each \gls{leo} satellite extends across a radius measuring hundreds of kilometers \cite{Choi2024}.
Notably, user devices follow non-uniform distribution patterns throughout the coverage area, forming distinct geographic clusters, as depicted in \figref{fig01}.
This distribution creates a scenario where users within individual clusters maintain close physical proximity, while substantial distances separate the clusters from one another \cite{Christopoulos2015}.
Consequently, this spatial configuration generates distinctive channel properties wherein intra-cluster user channels demonstrate substantially higher correlation compared to inter-cluster user channels \cite{Li2023a}.
This unique satellite channel characteristic can be exploited to enhance iterative MIMO detector convergence, forming the motivation for this research.

In this paper, a cluster-aware two-stage method is proposed for satellite MIMO detection that exploits the inherent clustering of users to significantly enhance iterative detector efficiency.
Stage 1 focuses on eliminating intra-cluster interference through small matrix inversions.
This approach is computationally efficient because each cluster contains only a limited number of users compared to the entire coverage area, making these small matrix inversions manageable.
In Stage 2, conventional iterative methods are applied to cancel the remaining inter-cluster interference.
Importantly, the inverted small matrices from Stage 1 serve as preconditioners that accelerate the convergence of these iterative methods.
Computer simulations demonstrate the advantages of the proposed method: under perfect \gls{csi} conditions, the approach achieves over $12$ times faster convergence compared to conventional iterative detectors.
Importantly, this advantage remains robust even in practical scenarios with channel estimation errors, where the method still maintains $9$ times faster convergence.
These significant convergence improvements establish the proposed approach as a compelling solution for efficient signal detection in next-generation satellite MIMO systems.

\section{System Model, Preliminaries and Problem Statement}\label{sec2}

\subsection{System Model}
Consider a satellite equipped with $M$ service antennas providing service to $N$ users over identical time-frequency resources.
The uplink satellite MIMO signal transmission can be mathematically expressed as follows
\begin{equation}\label{eqn01}
	\mathbf{y} = \mathbf{H} \mathbf{x} + \mathbf{v},
\end{equation}
where $\mathbf{y} \in \mathbb{C}^{M \times 1}$ denotes the received signal vector at the satellite, $\mathbf{H} \in \mathbb{C}^{M \times N}$ the random channel matrix, $\mathbf{x} \in \mathbb{C}^{N \times 1}$ the transmitted signals from the users, $\mathbf{v} \sim \mathcal{CN}(0,\sigma_{v}^{2}\mathbf{I})$ the \gls{awgn}, and $\mathbf{I}$ denotes an identity matrix with compatible dimensions; Each element of $\mathbf{x}$ is drawn from a finite alphabet set with equal probability, satisfying $\mathbb{E}\{\mathbf{x}\} = \mathbf{0}$ and $\mathbb{E}\{\mathbf{x}\mathbf{x}^{H}\} = \sigma_{x}^{2}\mathbf{I}$, where $\sigma_{x}^{2}$ denotes the transmission power.

Satellite MIMO channels typically follow Rician distribution due to the predominant \gls{los} propagation conditions \cite{3GPP2023}.
Denote $\mathbf{h}_{n}$ as the $n$-th column of $\mathbf{H}$, corresponding to the channel of the $n$-th user, it can be formulated as
\begin{equation}
	\mathbf{h}_{n} = \sqrt{P_{n}} \Bigg(\sqrt{\dfrac{K_{n}}{K_{n} + 1}} \mathbf{h}_{n}^{\textsc{los}} + \sqrt{\dfrac{1}{K_{n} + 1}} \mathbf{h}_{n}^{\textsc{nlos}} \Bigg),
\end{equation}
where $P_{n}$ represents the large-scale fading dependent on the link-budget calculation (see \cite{3GPP2023} or \secref{sec04}), and $K_{n}$ denotes the Rician $K$-factor following log-normal distribution, with its mean and variance depending on the elevation angle of the $n$-th user-to-satellite link \cite{3GPP2023};
The vector $\mathbf{h}_{n}^{\textsc{los}} \in \mathbb{C}^{M\times 1}$ denotes the phase of direct \gls{los} paths, with its $m$-th element defined as $h_{m,n}^{\textsc{los}} = \exp(-j\frac{2\pi}{\lambda}d_{m,n})$,  where $j$ represents the imaginary number unit, $\lambda$ the carrier wavelength, and $d_{m,n}$ the distance between the $n$-th user and $m$-th satellite antenna.
The vector $\mathbf{h}_{n}^{\textsc{nlos}} \in \mathbb{C}^{M\times 1}$ characterizes the small-scale fading behavior of the \gls{nlos} components with \gls{iid} complex Gaussian distribution, each element following $h_{m,n}^{\textsc{nlos}} \sim \mathcal{CN}(0, 1/M)$.

\subsection{Preliminaries} \label{sec02b}
The optimal linear MIMO detector for interference cancellation is \gls{lmmse}, which can be expressed as follows
\begin{equation} \label{eqn12040315}
	\widehat{\mathbf{x}} = \big(\mathbf{H}^{H} \mathbf{H} + \rho^{-1} \mathbf{I}\big)^{-1} \mathbf{H}^{H} \mathbf{y},
\end{equation}
where $\widehat{\mathbf{x}}$ denotes the estimated vector, and $\rho \triangleq \sigma_{x}^{2}/\sigma_{v}^{2}$ represents the \gls{snr}.
The problem \eqref{eqn12040315} can be rewritten into the following fixed point problem 
\begin{equation}\label{eqn04}
\mathbf{A} \widehat{\mathbf{x}} = \mathbf{b},
\end{equation}
where $\mathbf{A} \triangleq \mathbf{H}^{H} \mathbf{H} + \rho^{-1} \mathbf{I}$ is a Hermitian matrix, and $\mathbf{b} \triangleq \mathbf{H}^{H} \mathbf{y}$ represents the \gls{mf} vector.
It is evident that solving \eqref{eqn12040315} requires high computational complexity due to the matrix inversion operation.

Numerous iterative methods have been proposed to determine $\widehat{\mathbf{x}}$
while bypassing matrix inversion with quadratic-order complexity \cite{Albreem2019}.
One of the most classical iterative methods is \gls{ri}, which takes the form \cite{Tu2020}
\begin{equation} \label{eqn05}
	\widehat{\mathbf{x}}_{t+1} = \widehat{\mathbf{x}}_{t} + (\mathbf{b} - \mathbf{A}\widehat{\mathbf{x}}_{t}),
\end{equation}
where $t \geq 0$ denotes the iteration index. 
When $\mathbf{A}$ is ill-conditioning, \gls{ri} experiences slow convergence or even diverges.
For faster convergence, \gls{ms} based methods were proposed, which split the matrix $\mathbf{A}$ as follows
\begin{equation}\label{eqn06}
	\mathbf{A} = \mathbf{D} + \mathbf{L} + \mathbf{L}^H,
\end{equation}
where $\mathbf{D}$ is the diagonal part of $\mathbf{A}$, and $\mathbf{L}$ is the strict lower triangular part of $\mathbf{A}$.
MS-based methods utilize the inverse of part of $\mathbf{A}$ to accelerate convergence, as follows \cite{Liu2024c}
\begin{equation}\label{eqn07}
	\widehat{\mathbf{x}}_{t+1} = \widehat{\mathbf{x}}_{t} + \mathbf{M}^{-1} (\mathbf{b} - \mathbf{A} \widehat{\mathbf{x}}_{t}),
\end{equation}
where $\mathbf{M}$ is called preconditioning matrix.
Different MS-based methods define $\mathbf{M}$ differently, such as $\mathbf{M}_\text{Jac} = \mathbf{D}$ for Jacobi iteration, $\mathbf{M}_{\textsc{gs}} = \mathbf{D} + \mathbf{L}$ for \gls{gs} method \cite{Zhang2021}, and $\mathbf{M}_{\textsc{ssor}} = \mathbf{M}_{\textsc{gs}} \mathbf{D}^{-1}  \mathbf{M}_{\textsc{gs}}^H$ for \gls{ssor} \cite{Xie2016}.
Note that the the inversion of $\mathbf{M}$ in \eqref{eqn07} for GS and SSOR requires only quadratic complexity due to the triangular structure \cite{Liu2023e}.

\subsection{Problem Statement}
Current iterative methods for MIMO detection face slow convergence in satellite communications, as they overlook the distinctive channel properties illustrated in \figref{fig01}.
Specifically, the primary factor contributing to channel ill-conditioning in satellite MIMO systems is the high correlation among intra-cluster user channels, rather than inter-cluster channel correlation.
This property is confirmed by numerical results presented in \secref{sec04}.
However, existing iterative detection treat all user channels with equal consideration regardless of their spatial relationships.
This motivates the rest of this paper to leverage this unique satellite MIMO channel property.

\section{Efficient Two-Stage Method for Iterative Satellite MIMO Detection}
This section introduces the proposed two-stage satellite MIMO detection method exploiting clustered user distribution characteristics.
The algorithm operates sequentially: Stage 1 eliminates intra-cluster interference using small matrix inversions, followed by Stage 2 which iteratively cancels remaining inter-cluster interference.
Moreover, the section concludes with analysis of convergence properties and computational complexity, demonstrating theoretical advantages over current iterative detection methods.

\subsection{Stage 1: Intra-Cluster Interference Cancellation}
Consider a satellite system where users are organized into $C$ distinct clusters, with cluster $c$ contains $N_{c}$ users, such that $\sum_{c=0}^{C-1}N_{c} = N$.
Based on this clustering structure, the MIMO signal model in \eqref{eqn01} can be reformulated as follows
\begin{equation} \label{eqn12020222}
	\mathbf{y} = \mathbf{H}_{c} \mathbf{x}_{c} + \Bigg(\sum_{i\neq c}^{C} \mathbf{H}_{i} \mathbf{x}_{i}\Bigg) + \mathbf{v},
\end{equation}
where $\mathbf{H}_{c} \in \mathbb{C}^{M \times N_{c}}$  represents the channel matrix for users in cluster $c$, and $\mathbf{x}_{c} \in \mathbb{C}^{N_{c} \times 1}$ denotes their transmitted signal vector.
The following lemma establishes the theoretical foundation for intra-cluster interference cancellation:
\begin{lem}
	Suppose that the sub-channel matrices of different clusters are orthogonal to each other, i.e., $\mathbf{H}_{c}^{H} \mathbf{H}_{i} = \mathbf{0},\ \forall c\neq i$, the MSE of the estimated signal vector for the $c$-th cluster can be expressed as $\mathrm{MSE}_{c} = \|\mathbf{x}_{c} - \widetilde{\mathbf{x}}_{c}\|^{2}$, with $\widetilde{\mathbf{x}}_{c}$ denoting the estimated vector for users in cluster $c$.
	The optimal $\widetilde{\mathbf{x}}_{c}$ that minimizing $\mathrm{MSE}_{c}$ can be expressed as
	\begin{equation} \label{eqn080650316}
		\widetilde{\mathbf{x}}_{c} = \big(\mathbf{H}_{c}^{H} \mathbf{H}_{c} + \rho^{-1} \mathbf{I}\big)^{-1} \mathbf{H}_{c}^{H} \mathbf{y},
	\end{equation}
	which is equivalent to the LMMSE detector if cluster $c$ existed in isolation.
\end{lem}

\begin{IEEEproof}
	Given the orthogonality of channel columns between clusters, the LMMSE estimation within each cluster remains unaffected by interference from other clusters.
	Consequently, the LMMSE filter for each cluster can be determined solely by the intra-cluster channel and noise variance.
	The detailed proof is omitted in this conference version owing to space constraints.
\end{IEEEproof}

The intra-cluster interference cancellation described in \eqref{eqn080650316} can be efficiently performed in parallel across all clusters.
Define $\mathbf{\Phi}_{c} \triangleq \mathbf{H}_{c}^{H}\mathbf{H}_{c} + \rho^{-1} \mathbf{I}$, the intra-cluster cancellation process can be expressed in the following compact form
\begin{equation} \label{eqn09310316}
	\widetilde{\mathbf{x}} = \mathbf{\Phi}^{-1} \mathbf{b},
\end{equation}
where $\mathbf{\Phi} \triangleq \mathrm{diag}(\mathbf{\Phi}_{1}, \dots, \mathbf{\Phi}_{C})$ with $\mathrm{diag}(\cdot)$ representing the operation to form a block diagonal matrix using the input matrices.
Note that $\mathbf{b}$ is a concatenation of $\mathbf{b}_{c} \triangleq \mathbf{H}_{c}\mathbf{y}$ for all clusters, i.e., $\mathbf{b} = [\mathbf{b}_{1}^{H}, \cdots, \mathbf{b}_{C}^{H}]^{H}$, and similarly $\widetilde{\mathbf{x}} \in \mathbb{C}^{N \times 1}$ is the concatenation of all cluster estimates $\widetilde{\mathbf{x}}_{c}$, expressed as $\widetilde{\mathbf{x}} = [\widetilde{\mathbf{x}}_{1}^{H}, \cdots, \widetilde{\mathbf{x}}_{C}^{H}]^{H}$.
This formulation provides two significant computational advantages: \textit{1)} The matrix inversion of $\mathbf{\Phi}$ can be performed in parallel across all blocks, as each block corresponds to an independent cluster.
\textit{2)} The computational complexity is substantially reduced since matrix inversions are performed on smaller matrices of dimension $N_c$ instead of the full system matrix of dimension $N$.
A comprehensive complexity analysis of the proposed method will be provided in \secref{sec03c}.

While \eqref{eqn09310316} effectively mitigates intra-cluster interference, it does not address inter-cluster interference.
Though inter-cluster interference is comparatively weaker than intra-cluster interference, treating it merely as noise significantly degrades overall detection performance.
Therefore, eliminating inter-cluster interference is essential for achieving optimal LMMSE detection performance across the entire system.
This requirement motivates Stage 2 of the proposed method, which is detailed in the following subsection.

\subsection{Stage 2: Iterative Inter-Cluster Interference Cancellation}
In Stage 1, the block diagonal matrix $\mathbf{\Phi}$ was efficiently inverted through $C$ small matrix inversions (one per cluster).
This section builds upon that foundation to address the remaining inter-cluster interference.

The conventional iterative methods for solving the fixed point problem in \eqref{eqn04} are limited by the ill-conditioned nature of matrix $\mathbf{A}$. To overcome this limitation, the proposed approach utilizes $\mathbf{\Phi}^{-1}$
(already computed in Stage 1) as an effective preconditioner.
This is accomplished by multiplying $\mathbf{\Phi}^{-1}$ on both sides of \eqref{eqn04}, resulting in
\begin{equation} \label{eqn10460316}
	\mathbf{\Psi} \widehat{\mathbf{x}} = \widetilde{\mathbf{x}},
\end{equation}
where $\mathbf{\Psi} \triangleq \mathbf{\Phi}^{-1} \mathbf{A}$, and $\widetilde{\mathbf{x}}$ is the initial estimate obtained from Stage 1 as shown in \eqref{eqn09310316}.
It is important to note that \eqref{eqn10460316} and \eqref{eqn04} are equivalent and yield the same solution $\widehat{\mathbf{x}}$.
Therefore, standard iterative methods can be applied to solve \eqref{eqn10460316} as well.
A key advantage of the transformed system is that the diagonal component of $\mathbf{\Psi}$ is an identity matrix, allowing for the following decomposition
\begin{equation}
	\mathbf{\Psi} = \mathbf{I} + \mathbf{F} + \mathbf{F}^{H},
\end{equation}
where $\mathbf{F}$ represents the strict lower triangular part of $\mathbf{\Psi}$.
The iterative formulations of RI, GS, and SSOR methods for solving \eqref{eqn10460316} can be mathematically expressed as
\begin{equation}
	\widehat{\mathbf{x}}_{t+1} = \widehat{\mathbf{x}}_{t} + \mathbf{\Theta}^{-1} (\widetilde{\mathbf{x}} - \mathbf{\Psi} \widehat{\mathbf{x}}_{t}),
\end{equation}
where $\mathbf{\Theta}_{\textsc{ri}} = \mathbf{\Theta}_{\text{Jac}} = \mathbf{I}$ for RI and Jacobi iteration, $\mathbf{\Theta}_{\textsc{gs}} = \mathbf{I} + \mathbf{F}$ for \gls{gs}, and $\mathbf{\Theta}_{\textsc{ssor}} = \mathbf{\Theta}_{\textsc{gs}}  \mathbf{\Theta}_{\textsc{gs}}^{H}$ for \gls{ssor}.

The fundamental advantage of this two-stage approach is that the convergence rate of the iterative method is now governed by the condition number $\kappa(\mathbf{\Psi})$ rather than $\kappa(\mathbf{A})$.
A theoretical comparison of these condition numbers and their impact on convergence performance will be presented in the subsequent section.

\subsection{Convergence Rate Analysis}
This section provides a theoretical analysis demonstrating why the proposed two-stage method achieves significantly faster convergence compared to conventional approaches.

For iterative methods solving linear systems of the form $\mathbf{A}\widehat{\mathbf{x}}=\mathbf{b}$, the convergence rate is fundamentally determined by the condition number of matrix $\mathbf{A}$, defined as follows
\begin{equation}
	\kappa(\mathbf{A}) \triangleq \dfrac{\Lambda_{\max}(\mathbf{A})}{\Lambda_{\min}(\mathbf{A})},
\end{equation}
where $\Lambda_{\max}(\mathbf{\cdot})$ and $\Lambda_{\min}(\mathbf{\cdot})$ represent the maximum and minimum eigenvalues of the input matrix, respectively.
Conventional iterative methods exhibit asymptotic convergence rates inversely proportional to $\kappa(\mathbf{A})$, resulting in prohibitively slow convergence when $\kappa(\mathbf{A})$ is large \cite{Albreem2019}. 
The key advantage of the proposed method lies in transforming the original system into one with a significantly smaller condition number.
This is formally established by the following theorem:

\begin{thm} \label{thm01}
Let matrix $\mathbf{A}$ be decomposed as $\mathbf{A} = \mathbf{\Phi} + \mathbf{\Delta}$, where $\mathbf{\Delta}$ contains only the off-diagonal blocks of $\mathbf{A}$.
Suppose that: \textit{1)} the original channel matrix is ill-conditioned, i.e., $\kappa(\mathbf{A}) \gg 1$, and \textit{2)} the inter-cluster channel correlation is significantly smaller than the intra-cluster correlation, i.e., $\|\mathbf{\Phi}\| \gg \|\mathbf{\Delta}\| \approx 0$, then the following inequality holds
	\begin{equation}
		\kappa(\mathbf{\Psi}) \ll \kappa(\mathbf{A}).
	\end{equation}
\end{thm}

\begin{IEEEproof}
	By noting that $\mathbf{\Psi} = \mathbf{\Phi}^{-1} \mathbf{A} = \mathbf{I} + \mathbf{\Phi}^{-1}\mathbf{\Delta}$
	and applying Gershgorin's circle theorem, the eigenvalues of $\mathbf{\Psi}$ is bounded as follows \cite{Salas1999}
	\begin{equation}
		1 - \|\mathbf{\Phi}^{-1}\mathbf{\Delta}\| \leq \Lambda_{\min}(\mathbf{\Psi}) \leq \Lambda_{\max}(\mathbf{\Psi}) \leq 1 + \|\mathbf{\Phi}^{-1}\mathbf{\Delta}\|.
	\end{equation}
	Therefore, the condition number of $\mathbf{\Psi}$ is bounded as
	\begin{equation}
		\kappa(\mathbf{\Psi}) = \frac{\Lambda_{\max}(\mathbf{\Psi})}{\Lambda_{\min}(\mathbf{\Psi})} \leq \frac{1 +\|\mathbf{\Phi}^{-1}\mathbf{\Delta}\|}{1-\|\mathbf{\Phi}^{-1}\mathbf{\Delta}\|}.
	\end{equation}
	Since it is assumed that inter-cluster interference is weak ($\|\mathbf{\Delta}\| \approx 0$), we have $\|\mathbf{\Phi}^{-1}\mathbf{\Delta}\| \approx 0$. This implies that $\kappa(\mathbf{\Psi})$ approaches $1$.
	Given that $\kappa(\mathbf{A}) \gg 1$, we can conclude that $\kappa(\mathbf{\Psi}) \ll \kappa(\mathbf{A})$, which proves \thmref{thm01}.
\end{IEEEproof}

The result in \thmref{thm01} demonstrates a fundamental advantage of the proposed approach: iterative methods solving the preconditioned system in \eqref{eqn10460316} (Stage 2) converge substantially faster than conventional iterative methods directly applied to \eqref{eqn04}. 
However, this improved convergence comes at a cost: the proposed method requires additional computational operations, particularly the small matrix inversions in Stage 1.
The practical value of the two-stage approach depends on whether the accelerated convergence sufficiently outweighs this additional computational overhead.
The following subsection presents a comprehensive complexity analysis that quantifies the computational costs of the proposed two-stage method.

\subsection{Complexity Analysis} \label{sec03c}
This section presents a comprehensive complexity analysis of the proposed two-stage method, comparing it with conventional approaches.
A summary of all iterative algorithms is presented in \textsc{TABLE} \ref{tab01}.
For clarity of presentation, uniform cluster sizes are assumed where $N_{c} = \frac{N}{C}$ for all clusters.
Moreover, the complexity of addition and subtraction operations is assumed to be negligible in the table.

\begin{table}
	\caption{Complexity of Different MIMO Detection Methods}
	\label{tab01}
	\centering
	\renewcommand{\arraystretch}{0.4}
	\resizebox{0.49\textwidth}{!}{
		\begin{tabular}{ccc}
			\toprule
			Algorithms               & Matrix inverse &  Iterative complexity \\ 
			\midrule
			LMMSE    & $\mathcal{O}\big(N^{3}\big)$                        & $0$              \\ 
			\addlinespace
			RI (Original)    & 0               & $\mathcal{O}\big(T\big)\big(N^{2}+2N\big)$               \\ 
			\addlinespace
			GS (Original)    & $\mathcal{O}\big(N^{2}\big)$                  & $\mathcal{O}\big(T\big)\big(1.5N^{2}\big)$                 \\ 
			\addlinespace
			SSOR (Original)  & $\mathcal{O}\big(N^{2}\big)$                  & $\mathcal{O}\big(T\big)\big(2N^{2}+N\big)$               \\ 
			\addlinespace
			RI (Proposed)    & $\mathcal{O}\big(\frac{N^{3}}{C^{3}}\big)+ \frac{N^{3}}{C^{2}}+\frac{N^2}{C}$      & $\mathcal{O}\big(T\big)\big(N^{2}+2N\big)$               \\ 
			\addlinespace
			GS (Proposed)    & $\mathcal{O}\big(\frac{N^{3}}{C^{3}}+N^2\big)+ \frac{N^{3}}{C^{2}}+\frac{N^2}{C}$     & $\mathcal{O}\big(T\big)\big(1.5N^{2}\big)$                 \\ 
			\addlinespace
			SSOR (Proposed)  & $\mathcal{O}\big(\frac{N^{3}}{C^{3}}+N^2\big)+ \frac{N^{3}}{C^{2}}+\frac{N^2}{C}$     & $\mathcal{O}\big(T\big)\big(2N^{2}+N\big)$               \\ 
			\bottomrule
	\end{tabular}}
	\vspace{-1em}
\end{table}

Initially, computing $\mathbf{A}$ and $\mathbf{b}$ requires $MN^{2}$ and $MN$ operations, respectively.
They support full parallel implementation, enabling efficient computation on parallel processing hardware. 
Additionally, these computations are common prerequisites for all detection algorithms, including both LMMSE and iterative methods.
Hence, this complexity is ignored in the table.
For conventional LMMSE (or ZF) detectors, the primary computational bottleneck is the matrix inversion with complexity $\mathcal{O}(N^{3})$, where $\mathcal{O}(\cdot)$ denotes the time complexity that does not account for parallel computing benefits.
In contrast, Stage 1 of the proposed method requires $C$ small matrix inversions with dimension $\frac{N}{C}$, resulting in a total computational complexity of $\mathcal{O}\big(\frac{N^{3}}{C^{3}}\big)C$, which is also the complexity of calculating $\mathbf{\Phi}^{-1}$.

In Stage 2 of the proposed method, calculating the linear transformation from \eqref{eqn04} to \eqref{eqn10460316} is required.
Due to the block diagonal structure of $\mathbf{\Phi}$, this step requires calculating $\mathbf{\Psi}$ and $\widetilde{\mathbf{x}}$, with complexities $\frac{N^{3}}{C^{2}}$ and $\frac{N^2}{C}$, respectively.
These calculations consist only of matrix-matrix multiplications that support full parallel implementation, making them computationally efficient in practice.
In the iterative processes, computing $\mathbf{A}\widehat{\mathbf{x}}_{t}$ (or $\mathbf{\Psi}\widehat{\mathbf{x}}_{t}$) requires $N^{2}$ operations per iteration.
The \gls{gs} and \gls{ssor} methods involve triangular matrix inversions with serial complexity $\mathcal{O}(N^{2})$.
Moreover, due to the lower-triangular structure of $\mathbf{M}$ and $\mathbf{\Theta}$, multiplying $\mathbf{M}^{-1}$ or $\mathbf{\Theta}^{-1}$ with a vector has parallel complexity of $\frac{N^{2}}{2}$ and $N^{2}$, respectively.
The computational efficiency of the proposed method depends critically on its convergence enhancement capabilities. 
A comprehensive discussion of quantitative measurements of convergence acceleration is presented in \textit{Case Study 3} and \textit{Case Study 4} in the following section.

\section{Numerical and Simulation Results} \label{sec04}
This section presents comprehensive numerical and simulation results, aiming to demonstrate four key aspects:
\textit{1)} The intra-cluster channel correlation in  satellite MIMO systems is significantly higher than inter-cluster interference;
\textit{2)} The preconditioned matrix $\mathbf{\Psi}$ exhibits smaller condition number properties compared to the original system matrix $\mathbf{A}$;
\textit{3)} The proposed two-stage method provides substantially faster convergence rates compared to conventional iterative methods when perfect channel state information is available;
\textit{4)} The efficiency of the proposed method is maintained even in the presence of practical channel estimation errors.
We consider two scenarios of user configurations:

\textbf{Scenario 1:} There are $16$ users equally distributed among $4$ clusters, with each cluster containing $N_{c} = 4$ users.

\textbf{Scenario 2:} There are $64$ users equally distributed among $8$ clusters, with each cluster containing $N_{c} = 8$ users.

The link-budget calculations are conducted according to the 3GPP standard \cite[Section 6.1.3]{3GPP2023}.
Specifically, the satellite MIMO system operates at a frequency of $2.0$ GHz with a bandwidth of $2.0$ MHz.
The LEO satellite orbits at an altitude of $550$ km and the users transmit with $16$-$26$
dBm power.
The ambient and antenna temperatures of $260$ and $150$ Kelvin, respectively.
The link experiences a free space path loss of $153.55$ dB.
The satellite is equipped with a \gls{upa} comprising $48\times48$ elements, each providing an element gain of $3$ dBi.
The system achieves a G/T value of $14.96$ dB/Kelvin, and an overall SNR of $10$ to $20$ dB, depending on the transmission power.
The objective of this section sets the following four case studies:

\textit{Case Study 1:}
\begin{figure}[t]
	\centering
	\begin{minipage}[t]{0.49\textwidth}	
		\centering
		\includegraphics[width=5.46cm]{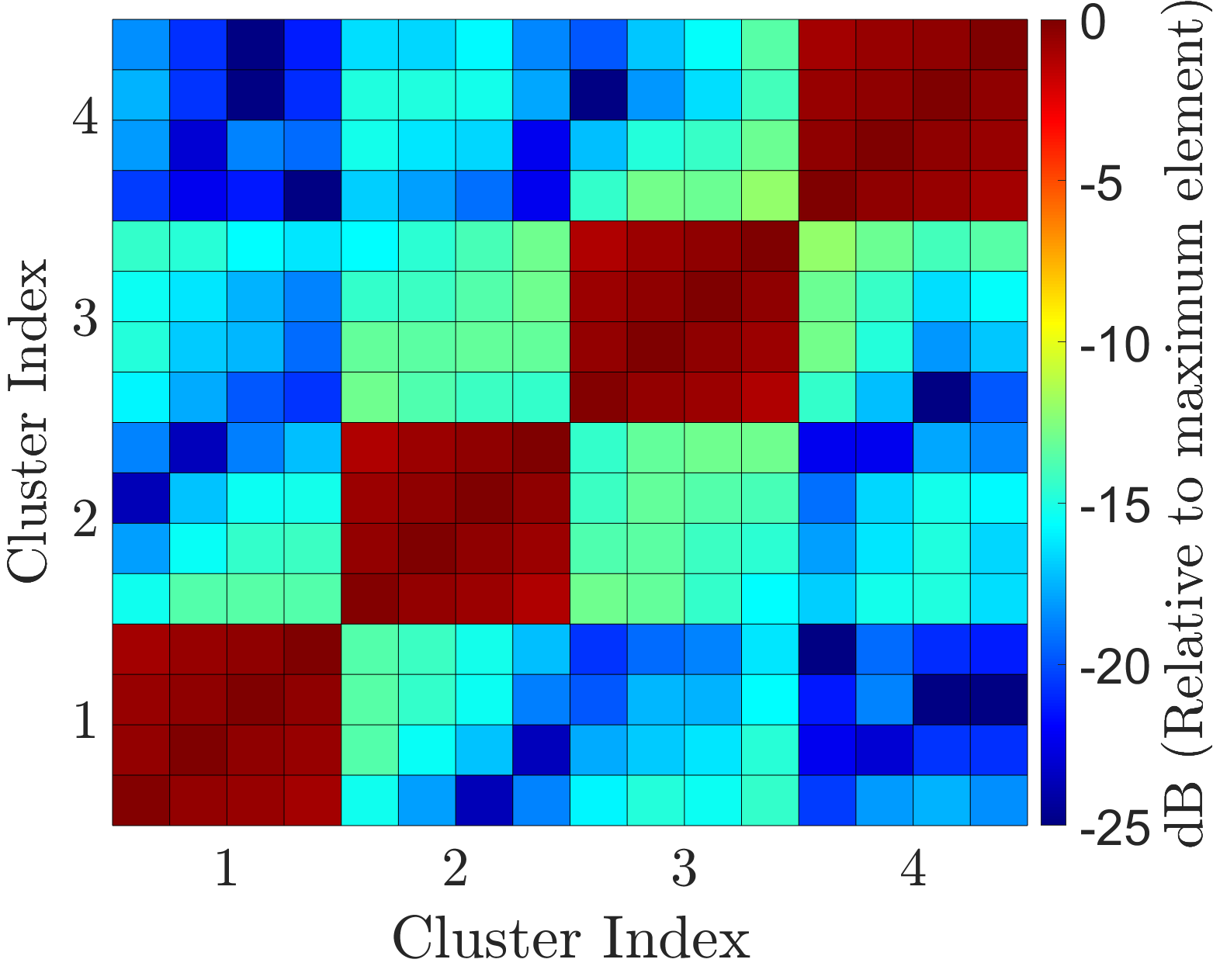}
	\end{minipage}
	\caption{\label{fig02} Heat map of instantaneous channel correlation matrix (i.e., $\mathbf{A}$) for \textbf{Scenario 1}. Red diagonal blocks show strong intra-cluster correlation, while blue-green off-diagonal regions indicate weak inter-cluster correlation.}
	\vspace{-1em}
\end{figure}
In this case study, the objective is to show the comparison between intra-cluster channel correlation and inter-cluster channel correlation.
Figure \ref{fig02} illustrates the amplitude of instantaneous matrix $\mathbf{A}$ for \textbf{Scenario 1}, normalized to the maximum element.
The heat map clearly reveals the block-diagonal structure of the channel correlation matrix, where each diagonal block represents intra-cluster correlation.
The diagonal blocks exhibit significantly higher correlation values (near $0$ dB), while the off-diagonal blocks representing inter-cluster correlation display much lower values, i.e., from $-10$ to $-25$ dB.
This confirms that the main factor contributing to the satellite channel ill-conditioning is the high intra-cluster channel correlation, while inter-cluster channel correlation is comparatively weaker.

\textit{Case Study 2:}
\begin{figure}[t]
	\centering
	\begin{minipage}[t]{0.49\textwidth}	
		\centering
		\includegraphics[width=6cm]{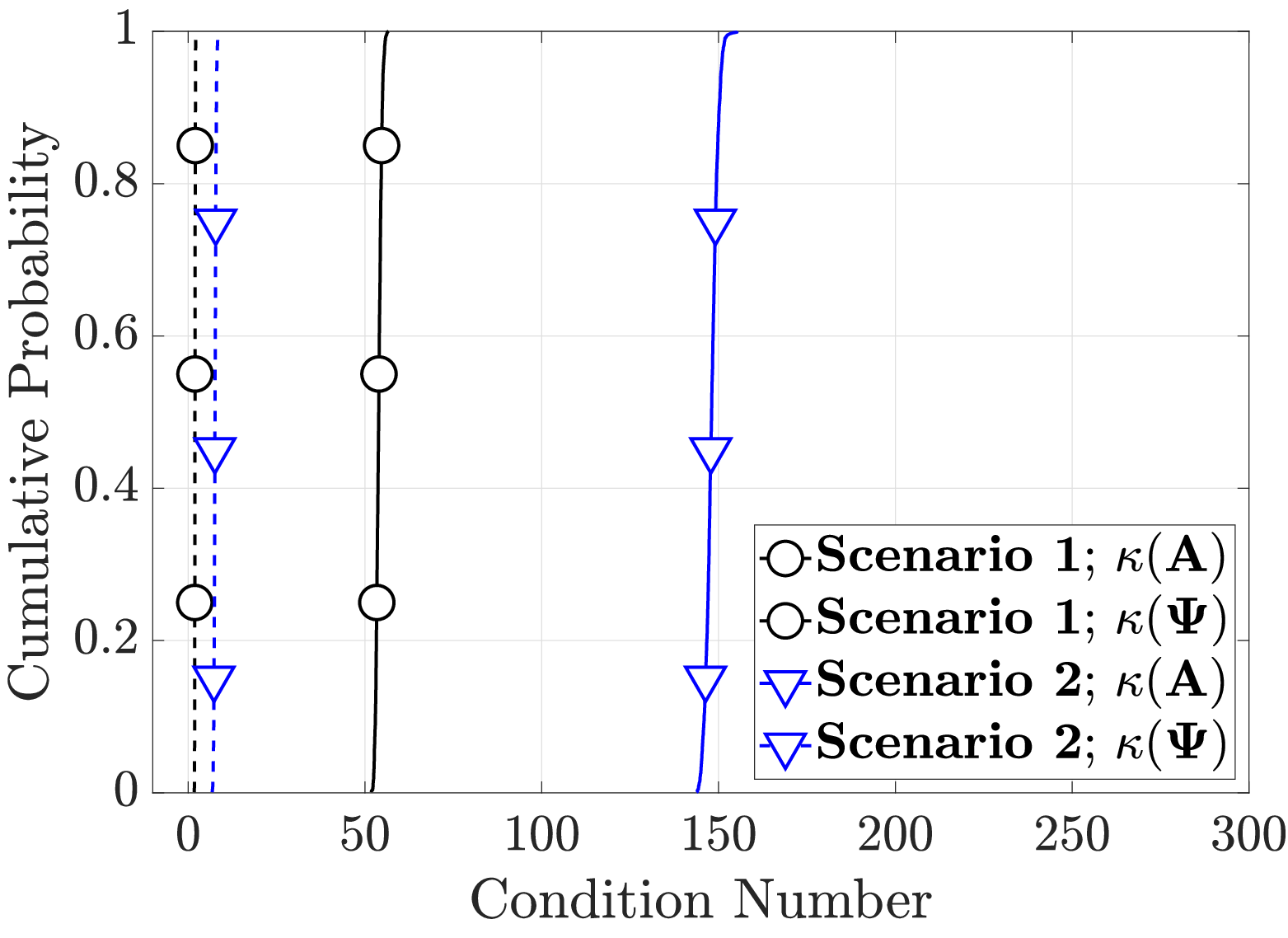}
	\end{minipage}
	\caption{\label{fig03} CDF of condition numbers for the original system matrix $\mathbf{A}$ and the preconditioned matrix $\mathbf{\Psi}$. It is obvious that $\mathbf{\Psi}$ exhibits significantly lower condition numbers, demonstrating the effectiveness of the proposed method.}
	\vspace{-1em}
\end{figure}
In this case study, the objective is to compare the condition numbers between $\mathbf{A}$ and $\mathbf{\Psi}$.
\figref{fig02} presents the \gls{cdf} of condition numbers for both the original system matrix $\mathbf{A}$ and the preconditioned matrix $\mathbf{\Psi}$ across two different scenarios.
For \textbf{Scenario 1}, the condition number of the original matrix $\mathbf{A}$ ranges around $50$ to $60$, while the preconditioned matrix $\mathbf{\Psi}$ exhibits condition numbers close to $1$ to $10$.
This represents approximately a $5$ to $6$ times reduction in the condition number.
The contrast is even more dramatic in \textbf{Scenario 2}, representing more than a more than $7$ times improvement.
This substantial reduction in condition numbers confirms the theoretical analysis presented in \thmref{thm01}. 
Also, this improvement can translates to faster convergence rates for iterative methods, which will be demonstrated in the following two case studies.

\textit{Case Study 3: }
\begin{figure}[t]
	\centering
	\subfigure[\textbf{Scenario 1}]{
		\begin{minipage}[t]{0.49\textwidth}	
			\label{fig01a}
			\centering
			\includegraphics[width=6.43cm]{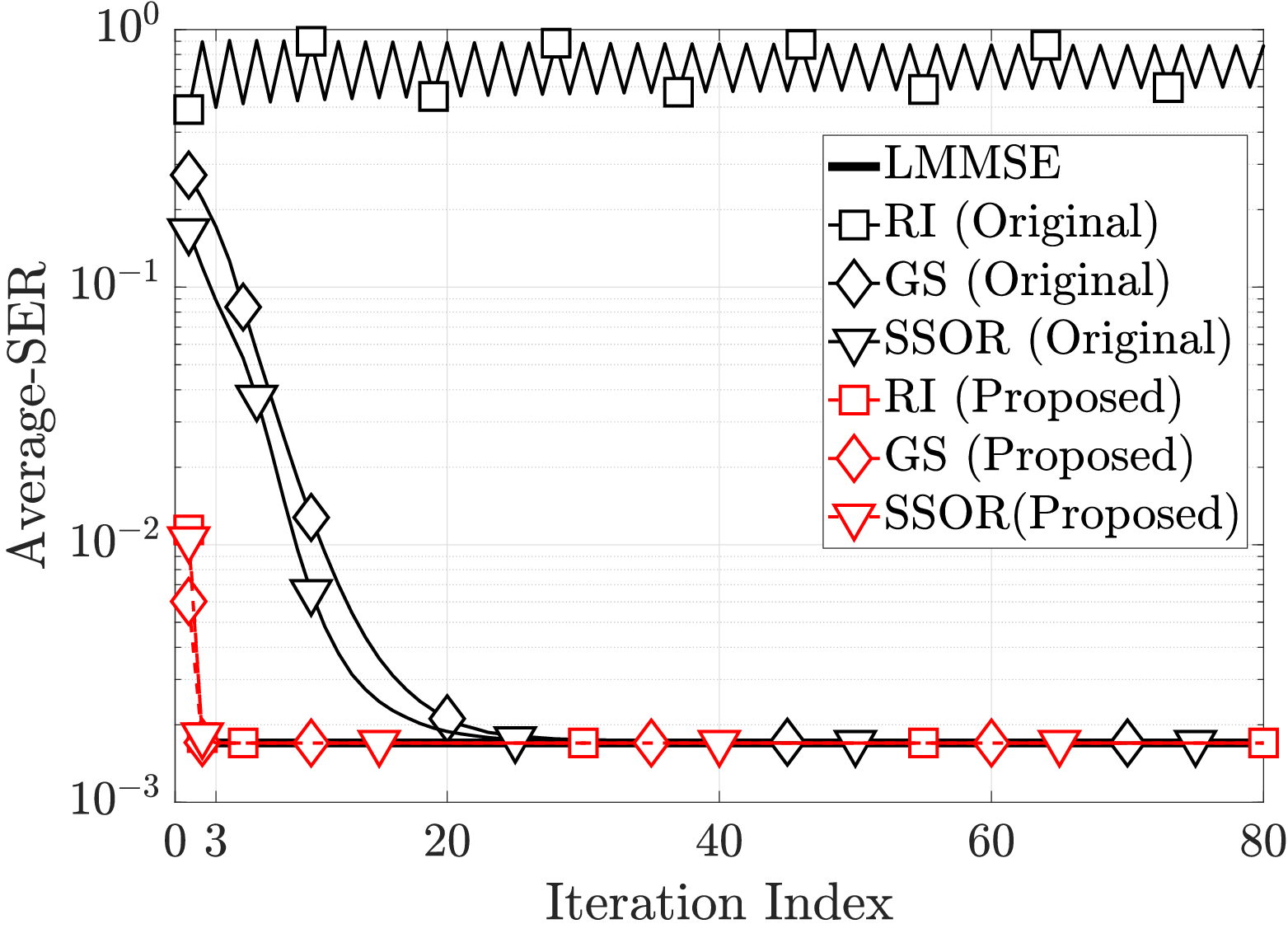}
			\vspace{0.5em}
	\end{minipage}}
	\subfigure[\textbf{Scenario 2}]{
		\begin{minipage}[t]{0.49\textwidth}
			\label{fig01b}
			\centering
			\includegraphics[width=6.43cm]{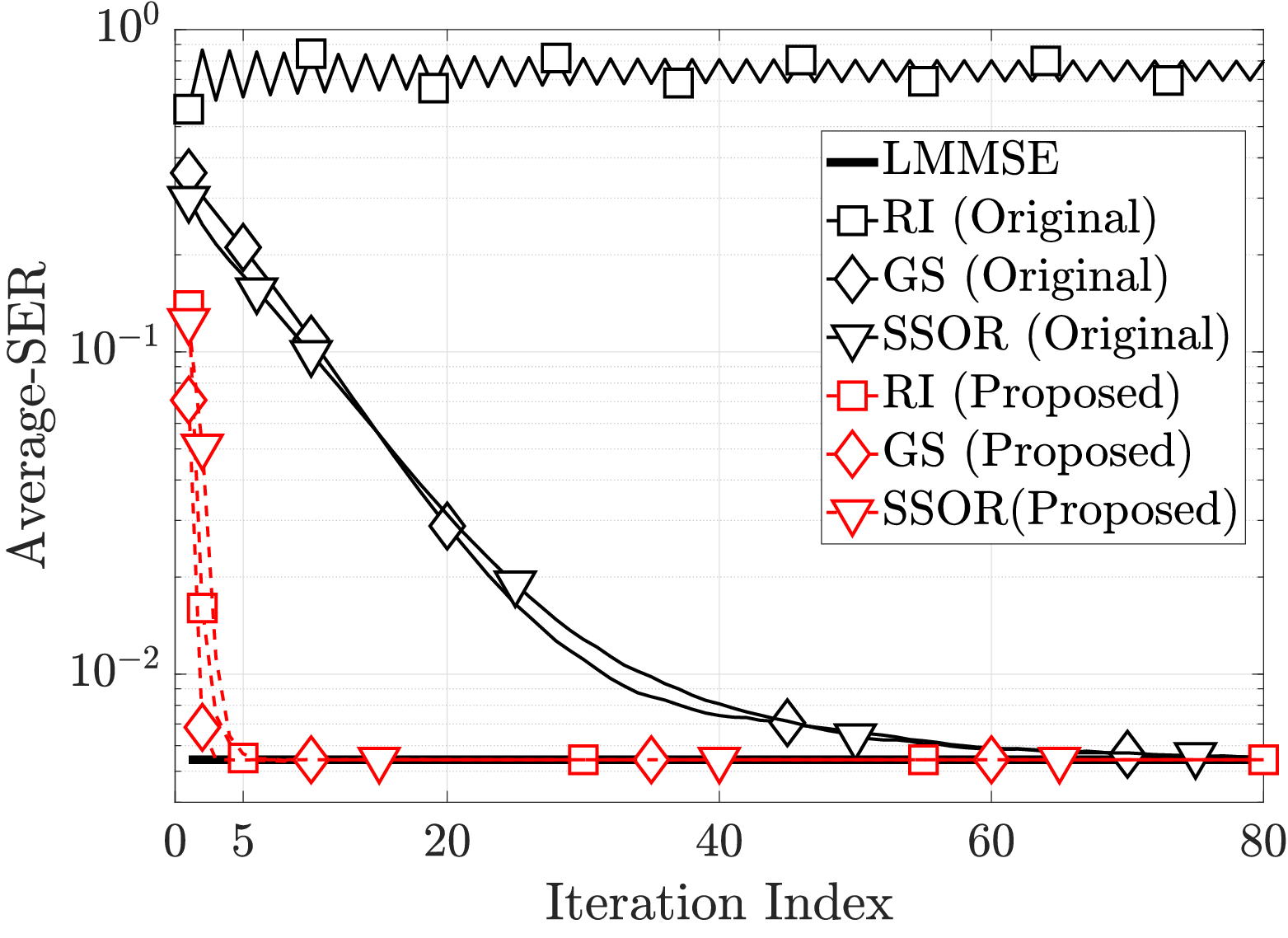}
			\vspace{0.5em}
	\end{minipage}}
	\caption{\label{fig04} Convergence comparison at SNR = $19$ dB shows the proposed method achieve LMMSE performance within few iterations ($2$ to $10$), while conventional approaches converge slowly or diverge.}
	\vspace{-1em}
\end{figure}
This case study compares the convergence rate of the proposed method against conventional approaches under perfect \gls{csi} conditions.
The \gls{ser} versus iteration index is illustrate in \figref{fig04}.
It can be found that conventional RI diverges completely while GS and SSOR methods converge extremely slowly, requiring over $20$ iterations in \textbf{Scenario 1} and more than $60$ iterations in the more challenging \textbf{Scenario 2}.
In contrast, the accelerated versions achieve LMMSE performance, i.e., \gls{ser}, within just $3$ iterations for \textbf{Scenario 1} and $5$ iterations for \textbf{Scenario 2}.
This demonstrates that the proposed method accelerates conventional methods by more than $12$ times in terms of convergence speed.
This significant acceleration of convergence is consistent with the condition number improvement demonstrated in \textit{ Case Study 2}.
These results confirm that the proposed method effectively addresses the ill-conditioning challenges inherent in satellite MIMO channels.

\textit{Case Study 4: }
\begin{figure*}[t]
	\centering
	\subfigure[\textbf{Scenario 1}]{
		\begin{minipage}[t]{0.49\textwidth}	
			\label{fig05a}
			\centering
			\includegraphics[width=6.4cm]{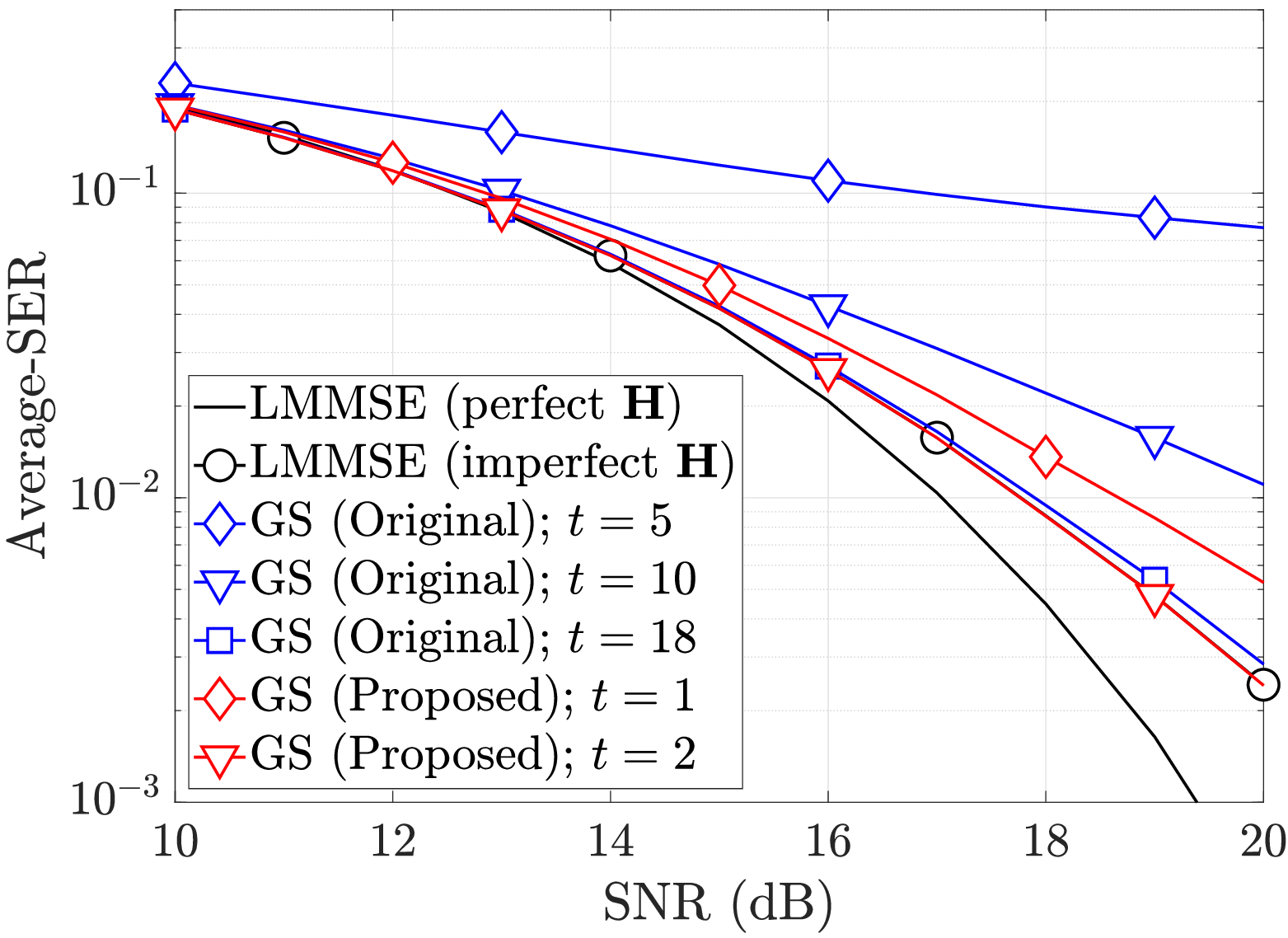}
			\vspace{0.5em}
	\end{minipage}}
	\subfigure[\textbf{Scenario 2}]{
		\begin{minipage}[t]{0.49\textwidth}
			\label{fig05b}
			\centering
			\includegraphics[width=6.4cm]{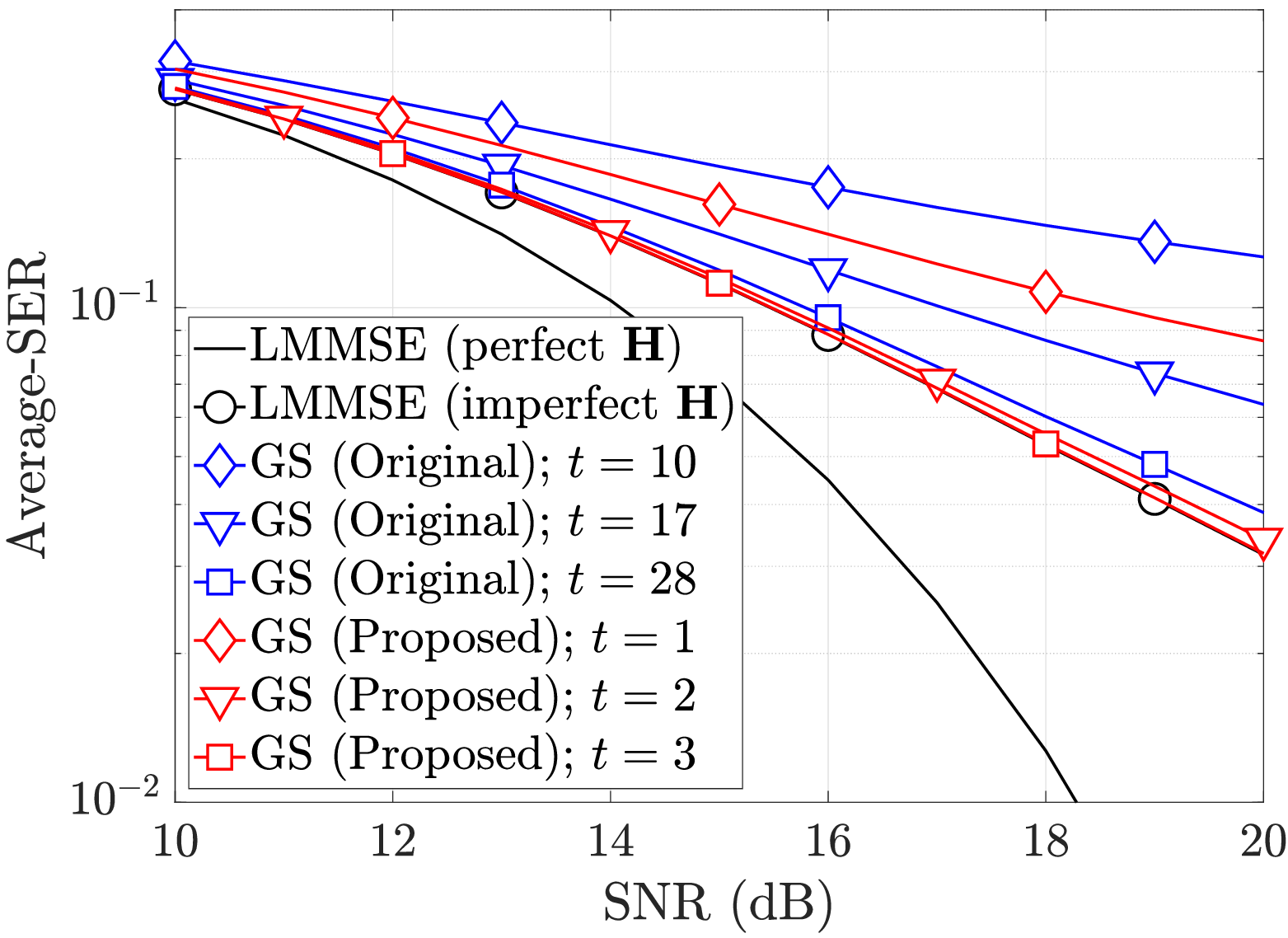}
			\vspace{0.5em}
	\end{minipage}}
	\caption{\label{fig05} SER performance comparison between conventional GS and the proposed method under imperfect CSI conditions. The proposed method achieves LMMSE performance within just $2$ or $3$ iterations, while conventional GS requires more than $18$ or $28$ iterations to reach comparable performance.}
	\vspace{-1em}
\end{figure*}
This case study evaluates the proposed method's robustness when confronted with practical channel estimation errors.
The simulation incorporates least square channel estimation with a normalized MSE of $10$ dB.
\figref{fig05} presents \gls{ser} versus \gls{snr} performance comparisons between conventional GS and the proposed method.
Critically, even under these challenging conditions, the proposed method maintains its substantial convergence advantage.
While conventional GS requires $18$ iterations to achieve LMMSE performance in \textbf{Scenario 1} and $28$ iterations in \textbf{Scenario 2}, the proposed method accomplishes equivalent performance with merely $2$ and $3$ iterations, respectively.
This represents approximately $9$ times faster convergence with channel imperfections.
Similar convergence advantages were observed with RI and SSOR methods, though these results are omitted due to space constraints.
This demonstrated resilience to channel estimation errors confirms the proposed method's practical viability for real-world satellite communication systems, where such estimation errors are unavoidable.

\section{Conclusion}
This paper proposed a cluster-aware two-stage detection method for satellite MIMO communications.
The approach leverages the clustered structure of satellite MIMO channels, where intra-cluster correlations significantly exceed inter-cluster correlations.
By separating intra-cluster interference cancellation from preconditioned inter-cluster interference elimination, the method achieves remarkable computational efficiency.
Theoretical analysis established that the proposed method substantially improves the iterative matrix condition number, enabling significantly faster convergence of iterative detection algorithms.
Computer simulations demonstrated that the proposed method achieves over $12$ times faster convergence under ideal channel conditions, while maintaining $9$ times faster convergence when considering channel estimation errors.
These advantages make the method particularly suitable for the next-generation satellite communications.

\appendices
\section*{Acknowledgment}
This work was funded by the 5G and 6G Innovation Centre, University of Surrey.

\ifCLASSOPTIONcaptionsoff
\newpage
\fi

\bibliographystyle{IEEEtran}
\bibliography{../IEEEabrv,../thesis_list}
\end{document}